\documentstyle[12pt]{article}
\begin{document}
\def\diagram#1{{\normallineskip=8pt\normalbaselineskip=0pt \matrix{#1}}}
\def\lrarr#1#2{\smash{\mathop{\raise.5ex\hbox to
.5in{\leftarrowfill} \hskip-.5in\raise-.5ex\hbox to
.5in{\rightarrowfill}} \limits^{\scriptstyle#1}_{\scriptstyle#2}}}
\def\rlarr#1#2{\smash{\mathop{\raise.5ex\hbox to
.5in{\rightarrowfill} \hskip-.5in\raise-.5ex\hbox to
.5in{\leftarrowfill}} \limits^{\scriptstyle#1}_{\scriptstyle#2}}}
\def\duarr#1#2{\llap{$\scriptstyle #1$}\downarrow \uparrow \vcenter to
.5in{}\rlap{$\scriptstyle #2$}}
\def\udarr#1#2{\llap{$\scriptstyle #1$}\uparrow \downarrow \vcenter to
.5in{}\rlap{$\scriptstyle #2$}}
\def\uarr#1#2{\llap{$\scriptstyle #1$}\uparrow \vcenter to
.5in{}\rlap{$\scriptstyle #2$}}
\def\rrarr#1#2{\smash{\mathop{\raise.5ex\hbox to
.5in{\rightarrowfill} \hskip-.5in\raise-.5ex\hbox to
.5in{\rightarrowfill}} \limits^{\scriptstyle#1}_{\scriptstyle#2}}}
\def\larr#1#2{\smash{\mathop{\hbox to
.5in{\leftarrowfill}}\limits^{\scriptstyle#1}_{\scriptstyle#2}}}

\def\harr#1#2{\smash{\mathop{\hbox to
.5in{\rightarrowfill}}\limits^{\scriptstyle#1}_{\scriptstyle#2}}}
\def\varr#1#2{\llap{$\scriptstyle #1$}\downarrow \vcenter to
.5in{}\rlap{$\scriptstyle #2$}}

\vglue 3cm
\begin{center}
{\large\sc{PRINCIPAL BUNDLES, CONNECTIONS AND BRST COHOMOLOGY}}\\[8mm]

{\sc H. Garc\' \i a-Compe\' an, J.M. L\' opez-Romero, M.A. Rodr\'
\i guez-Segura and  M. Socolovsky}

\vskip2cm

{\sc Abstract}

\end{center}

\noindent
{\small We review the elementary theory of gauge fields and the
Becchi-Rouet-Stora-
Tyutin symmetry in the context of differential geometry. We emphasize the
topological nature of this symmetry and discuss a double Chevalley-Eilenberg
complex for it.}\\

\vskip1cm

\section{Introduction}
 From their appearence {\it gauge theories} [1] have had a large influence on
both physics and mathematics. On the physical side one can date back to
Maxwell-Faraday (MF) abelian gauge theory (AGT) unifying electric and magnetic
phenomena (1860-70); the Einstein theory of general relativity (GR) (1915)
describing the gravitational force; first attemps of Weyl (1919), and Kaluza
and Klein (1919, 1926) to unify electromagnetism with gravity; the birth of
non abelian gauge theories (NAGT) in 1954 with the seminal work of Yang and
Mills (YM) which together with ideas from solid state physics (basically that
of spontaneous symmetry breaking) led to the SU(2)$\times $U(1)
Glashow - Weinberg - Salam (1961, 1967, 1968) electroweak (EW) theory unifying
the electromagnetic and the weak nuclear forces. (Notice that
from the geometrical point of view the EW theory involves just two
spheres: $S^3$ and $S^1$.) We should stress here that till now we have only
mentioned the {\it classical} parts of the corresponding theories; {\it
quantum} electrodynamics (QED) and quantum NAGT's were proved to be
renormalizable {\it i.e.} capable of absorbing infinite quantities appearing
in perturbation theory through the re-normalization of masses and coupling
(interaction) constants, after the contribution of a large number of
physicists, among others Dirac, Feynman, Dyson,
Tomonaga, Schwinger, Salam and Ward for the case of QED (1930-50) and
Faddeev, Popov, t'Hooft and Veltman for quantum NAGT's (1967, 1971-72). The
theory of the strong nuclear force based on the group $SU(3)$ (unfortunately
not a sphere!), quantum chromodynamics (QCD) found its place in the present
context after the works of Politzer, Weinberg, Gross and Wilczek (1973), thus
leading to the succesful  $SU(3)\times SU(2)\times U(1)$ standard model (SM)
for the electronuclear (EN) interactions. Unfortunately it does not exist at
present a renormalizable theory of quantum gravity (QG); following Hawking [2]
we might say that perhaps the greatest problem of theoretical physics in the
last quarter of our century is the conciliation of quantum mechanics and
general relativity. In this field there are at least two approaches: one is
that of Ashtekar and co-workers [3] who mantain general relativity as the
classical limit but construct the quantum theory after redefining the
fundamental variables, so the hope exists of constructing a theory of quantum
general relativity (QGR); the other is that of string theory (ST) (1970, 1974,
1984) [4] which represents a deep departure from the usual description of
elementary particles since at the roots of the theory is the idea that the
fundamental objects in Nature are not point-like but {\it string-like i.e.}
extended objects (even at the classical level!) though extremely small (of the
order of $10^{-35}m$) so that GR is modified at short distances and therefore
a theory of quantum gravity should not be QGR but what we might call quantum
string gravity (QSG); obviously GR as a classical theory is recovered from ST
in the large distance limit.

On the mathematical side the list of applications of gauge theory is much
shorter, but however of great importance: it consists in the application done
by Donaldson (1983) [5] of the theory of moduli spaces of instantons (self-
dual and anti-self-dual solutions to the classical YM equations based on the
group $SU(2)$) to the problem of classification of closed orientable 1-
connected differentiable 4-manifolds, and its relation to the same problem but
for topological manifolds previously considered by Freedman (1982) [6].

 From the physical point of view the basic idea behind gauge theory [1] is to
extend a {\it global} symmetry of a Lagrangian describing a particular set of
free (non-interacting) fields to a {\it local} symmetry {\it i.e.} one in
which the symmetry transformations of the fields can be done in an independent
way at each space-time point. One of the  most beautiful and important results
of this procedure is the appearance of {\it physical interactions} (couplings)
among the originally non-interacting fields. For example starting from the
Dirac Lagrangian $ {\cal L} = \bar \psi (\gamma ^\mu \partial _\mu +
m)\psi $ describing free electrons and positrons ( $\psi $ and $\bar
\psi $ fields) which has a global $U(1)$ symmetry $ \psi \to \psi ' =
e^{i\alpha} \psi $, $ \bar \psi \to \bar \psi ' = \bar \psi e^{-
i\alpha} $ ($ \alpha =$ const. $ \in {\bf R}$ and $ \gamma ^\mu $ : Dirac
matrices, $ \mu = 0,1,2,3 $), the introduction of the electromagnetic
(photon) field (gauge potential) through the replacement $ \partial
_\mu \to D_\mu := \partial _\mu + A_\mu $, with $ A_\mu $ transforming
as $ A_\mu (x) \to A'_\mu (x) = A_\mu (x) + i \partial _\mu \alpha (x)
$ (notice the locality $ \alpha = \alpha (x) $!) leads to the
interaction between the electron-positron field and the photon field
{\it i.e.} to the term $ \bar \psi \gamma^ \mu A_ \mu \psi $ in the QED
Lagrangian, $ {\cal L}_ {QED} = \bar \psi ( \gamma ^\mu D_\mu + m) \psi $.
Also, while global symmetries lead to conserved quantities (like electric
charge), local symmetries lead to Ward-Takahashi-Slavnov-Taylor identities
among Green's functions that is crucial for the proof of renormalizability.

Geometrically, classical gauge theories are {\it theories of connections on
principal fiber bundles} (p.f.b.'s), and related concepts in associated
bundles (like sections and covariant derivatives). In this framework
the gauge potentials of physics are local pull-backs on the base space
(typically a space-time) of connections which are differential 1-forms
globally defined on the total space of the bundle satisfying a set of
suitable conditions and with values in the Lie algebra of the symmetry
group of the theory (the fiber of the bundle), e.g. $U(1)$ in the
case of QED, $ SU(2) \times U(1) $ for EW theory, $ SU(3) $ for QCD,
$SO(3,1)$ for GR, etc. It is interesting to mention that even a concept
like spontaneous symmetry breaking has been incorporated into the bundle
language, see e.g. ref. [7]. The introduction of p.f.b.'s
inmediately leads us to consider problems in infinite dimensional
geometry, e.g. that of the gauge group of the bundle and its Lie
algebra; as we shall see in this article it is precisely the cohomology
of this algebra which leads to the extension (BRST symmetry) into the
quantum domain of the classical gauge symmetry.

The purpose of the present article is to review some basic ideas
involved in the theory of gauge fields from the geometrical viewpoint.
The material presented here is not original, except possibly the idea
of extending the usual BRST complex into a double complex
which could be naturally studied through the use of spectral sequences.
In section 2 we define principal and associated fiber
bundles, discuss the gauge group and its Lie algebra (without entering
into the difficulties of the relevant analysis), comment about the idea
of classification of p.f.b.'s, and discuss vector spaces of sections of
suitable vector bundles which are relevant for the next section. In
section 3 the concept of connection is introduced in its four different
variants, together with the definitions of: curvature; gauge
transformation of a connection and its  relation to the local
transformations in physics, which can then be understood as changes of
local trivializations of the bundle; covariant derivative and parallel
transport in associated bundles; and the YM function. Finally we
discuss the total space of connections and its quotient by the gauge
group. In section 4, we briefly discuss the Bonora and Cotta-Ramusino
[19] geometrical interpretation of the (quantum) BRST symmetry and cohomology
[20] of gauge theories, as the Chevalley-Eilenberg cohomology of the Lie
algebra of the gauge group with coefficients in the space of functions on the
space of connections. The definition of the relevant coboundary operator only
depends on the principal fiber bundle in question and is independent of any
particular connection (with the possible exception of a base point), thus
suggesting a deep relation between the topology of fiber bundles and quantum
mechanics. The definition of the usual BRST cochain complex allows an
inmediate generalization into a doble complex whose total cohomology could in
principle be computed through the use of spectral sequences.

\section{Principal and associated bundles}

In this section we shall present as much information as we need about a
principal fiber bundle (or an equivalence class of principal fiber bundles)
without using the notion of a connection (gauge field), which is an additional
structure that we can impose on a principal bundle and which basically allows
to define the concepts of parallel transport in the total bundle space and of
covariant derivatives of sections (matter fields) in associated bundles
(section 3), which physically lead to the concept of coupling (interaction)
between the matter and the gauge fields. As we shall see in section 4 the
concept of BRST cohomology in its algebraic formulation is a property of the
bundle itself and does not involve any particular connection, except for the
choice of a base point. In fact it depends on the total space of connections
which is a natural object associated with the bundle. In this sense we
can argue that the BRST cohomology is a property of the "space"  where the
connections live.

A smooth {\it principal fiber bundle} (p.f.b.) is a sextet $\xi =
(P,B,\pi ,G,$   ${\cal U}, \psi )$ where $P$ (total space) and $B$ (base
space) are respectively $s+r$- and $s$-dimensional differentiable manifolds,
$P\buildrel \pi \over \longrightarrow B$ (projection) is a smooth surjective
function, $G$ is an $r$-dimensional Lie group (structure group) which acts
freely and smoothly on $P$ through $P\times G\buildrel \psi \over
\longrightarrow P$, $\psi
(p,g)=\psi_g(p)$ $(=pg)$, $\psi ^{-1}_g=\psi _{g^{-1}}$, and transitively on
fibers $G_b=\pi ^{-1}(\{ b\} )$, $b\in B$  {\it i.e.} for all $p,q\in G_b$
there exists $g\in G$ such that $q=pg$ (since for any $b\in B$, $G_b$ is
diffeomorphic to G, one says that G is {\it the} fiber of the bundle); ${\cal
U}$ is an atlas on $\xi $ {\it i.e.} ${\cal U}=\{ (U_\alpha ,\phi
_\alpha)\}_{\alpha \in J}$ with open $U_\alpha \subset B$ and $P_\alpha = \pi
^{-1}(U _\alpha )\buildrel {\phi _\alpha } \over \longrightarrow U _\alpha
\times G$, $\phi _\alpha (p)=(\pi (p),\gamma _\alpha (p))$ a diffeomorphism
satisfying the condition $\pi _1\circ \phi _\alpha =\pi
_\alpha$ $(\pi _\alpha =\pi |_{P_\alpha })$; $\gamma _\alpha :P_\alpha
\longrightarrow G$ is smooth and satisfies $\gamma _\alpha (pg)=\gamma _\alpha
(p)g.$ Two p.f.b.'s $\xi =(P,B,\pi , G,{\cal U},\psi )$ y $\xi '=(P',B, \pi
',G,{\cal U}',\psi ')$ are equivalent ($\xi \cong \xi '$) if there exist smooth
maps $\alpha :P\to P'$, $\beta :P'\to P$ such that: 1) $\pi '\circ \alpha =\pi
, \pi \circ \beta =\pi '$, 2) $\alpha ,\beta $ are $G$-equivariant , 3) $\beta
\circ \alpha =1_P$, $\alpha \circ \beta =1_{P'}$. (For simplicity, in the
following we shall use the notation $\xi :G \longrightarrow P \buildrel
\pi\over \longrightarrow B$, saying that $\xi $ is a p.f.b. on $B$ or a $G$-
bundle on $B$.)

Given $\xi '$ and $\xi $ p.f.b.'s, a {\it bundle map} $\xi '\longrightarrow
\xi $ is a triple $(\alpha, \beta , h)$ where $P'\buildrel \alpha \over
\longrightarrow P$ and $B'\buildrel \beta \over \longrightarrow B$ are smooth
functions and $G'\buildrel h\over \longrightarrow G$ is a Lie group
homomorphism such that $\psi \circ (\alpha \times h)= \alpha \circ \psi '$ and
$\pi \circ \alpha =\beta \circ \pi '$; notice that $\alpha $ induces $\beta $:
if $b'\in B'$, there exists $ p'\in P'$ such that $b'=\pi '(p')$, then $\beta
(b')=\beta \circ \pi '(p')=\pi (\alpha (p'))$; the composition of bundle maps
is given by $(\alpha ',\beta ', h')\circ (\alpha ,\beta ,h)=(\alpha '\circ
\alpha , \beta '\circ \beta ,h'\circ h).$ If $\alpha $ and $\beta $ are
diffeomorphisms and $h$ is a Lie group isomorphism, then $\xi $ and $\xi '$
are isomorphic p.f.b.'s. In particular for $P'=P$, $B'=B$ and $G'=G$ the set
${\cal G}={\cal G}(\xi )=\{ (\alpha ,id_B,id_G)\} _{\alpha \in Diff(P)}$ is an
infinite dimensional Lie group $ [8] $ called the {\it group of vertical
automorphisms of $\xi $} or the {\it gauge group of $\xi $} $ [9] $. Each
element of ${\cal G}$ is represented by the following commutative diagram:

$$\diagram{
P\times G&\harr{\alpha \times id_G}{}&P\times G\cr
\varr{\psi }{}&&\varr{}{\psi }\cr
P&\harr{\alpha }{}&P\cr
\varr{\pi }{}&&\varr{}{\pi }\cr
B&\harr{id_B}{}&B\cr
}$$

We can give a second version of the gauge group of $\xi $:  let $\Gamma
_{eq}(P,G)=\{ \gamma :P\longrightarrow G$ smooth, $\gamma (pg)=g^{-1}\gamma
(p)g\} $ with the composition law $\gamma \cdot \gamma '(p)=\gamma (p)\gamma
'(p)$; there is a group isomorphism ${\cal G}(\xi )\buildrel \Sigma \over
\longrightarrow \Gamma _{eq}(P,G)$ given by: $\Sigma (\alpha )(p)$ is such that
$p\Sigma (\alpha )(p)=\alpha (p)$. A third version of ${\cal G}(\xi )$ will be
discussed after the concept of associated bundle is presented.

A {\it section} of a p.f.b. $\xi $ is a smooth function $B\buildrel s\over
\longrightarrow P$ such that $\pi \circ s=id_B$. One can prove that $\xi $ is
{\it trivial}, {\it i.e.} there exists a $G$-equivariant diffeomorphism
$P\buildrel \phi \over \longrightarrow B\times G$ such that $\pi _1\circ \phi
=\pi $ if and only if $\xi $ has a section (given the section the global
trivialization is $\phi (p)=(b,g)$ with $b=\pi (p)$ and $g\in G$ such that
$p=s(b)g$; given $\phi $ the section is $s(b)=\phi ^{-1}(b,e)$ with $e$ the
identity in $G$). A {\it local section} in $\xi $ is a smooth function
$s_\alpha :U_\alpha \to P_\alpha $ such that $\pi _\alpha \circ s_\alpha
=id_{U_\alpha }$; given the atlas ${\cal U}$ of a p.f.b. a canonical set of
local sections is $\sigma _\alpha (b)=\phi _\alpha ^{-1}(b,e)$.

If $\xi $ is a p.f.b. on $B$ and $f:B'\to B$ is a smooth function, then the
{\it pull-back bundle} $f^*(\xi )$ on $B'$ has structure group $G$ and total
space $P'=f^*(P)=\{ (b',p)|f(b')=\pi (p)\} \subset B'\times P$. Clearly
$(p_2,f,id_G)$ is a bundle map $f^*(\xi )\to \xi $. A well known construction
due to Milnor [10] says that for any Lie group $G$ (in fact the construction
extends to any topological group $G$) there exists a {\it universal principal
bundle}  $\xi G:G\to EG\buildrel \pi \over \to BG$ (unique up to homotopy
type) such that for any $G$-bundle $\xi '$ on $B'$ there exists a smooth
function (unique up to homotopy) $B'\buildrel f\over \to BG$ with $\xi
'\cong f^*(\xi G)$; in other words, if ${\cal B}_B(G)$ denotes the equivalence
classes of p.f.b.'s on $B$ with structure group $G$, and $[B,BG]$ is the set
of homotopy classes of maps from $B$ to $BG$, then
${\cal B}_B(G)\cong  [B,BG]$; in particular for $G = {\bf Z}_2\cong  S^0$,
$U(1)\cong  S^1$ and $SU(2)\buildrel ~\over \cong S^3$ one respectively has
$BS^0\cong  {\bf R}P^\infty $, $BS^1\cong {\bf C}P^\infty $ and $BS^3\cong {\bf
H}P^\infty $; and ${\cal B}_{S^1}(S^0)\cong \pi _1({\bf R}P^\infty )\cong \pi
_0(S^0)\cong {\bf Z}_2$, ${\cal B}_{S^2}(S^1)\cong \pi _2({\bf C}P^\infty
)\cong \pi _1(S^1)\cong {\bf Z}$ and ${\cal B}_{S^4}(S^3) \cong \pi _4({\bf
H}P^\infty)\cong \pi _3(S^3)\cong {\bf Z}$. In physical applications,
$SU(2)-bundles$ on $S^4={\bf
R}^4\cup \{ \infty \} $ and $U(1)-bundles$ on $S^2={\bf R}^2\cup\{ \infty \} $
are important examples, the choice of an equivalence class of such bundles
being equivalent to the choice of an integer number, in physical terms the
{\it winding number} also called the {\it instanton} or {\it monopole} number
for the $SU(2)$ and $U(1)$ cases respectively.

Let $X$ be a differentiable manifold and $G\times X\buildrel \mu \over
\to X$, $\mu (g,x)=\mu _g(x)(=g\cdot x)$, $\mu _g^{-
1}=\mu _{g^-1}$ a smooth left action of $G$ on $X$. Together with the p.f.b.
$\xi $
this action induces the {\it associated fiber bundle} $\xi _X:X - P\times
_GX\buildrel {\pi _X}\over \to B$ with fiber $X$, total space $P_X=P\times
_GX=\{ \langle p,x\rangle \} _{(p,x)\in P\times  X}$, $\langle p,x\rangle =\{
(pg,g^{-1}\cdot x)\} _{g\in G}$, base space $B$, projection $\pi _X(\langle
p,x\rangle )=\pi (p)$, and local triviality condition given by $P_\alpha
\times  _GX\buildrel {\Phi _\alpha }\over \to U_\alpha \times  X$, $\Phi
_\alpha (\langle p,x\rangle )=(\pi (p), \gamma _\alpha (p)\cdot x)$ in a given
 atlas ${\cal U}$ of $\xi $. If $X=V$ is a real (complex) n-dimensional vector
space then $\xi_V$ is called a real (complex) vector bundle of rank n. There
is a bijection between the set of equivariant functions $\Gamma _{eq}
(P,X)=\lbrace \gamma : P \to X$ smooth, $\gamma (pg)=g^{-1} \cdot \gamma
(p)\rbrace$ and the set of sections of $\xi_X,\Gamma (\xi_X)$: in fact $\gamma
\in \Gamma_{eq}(P,X)$ induces $s_\gamma \in \Gamma( \xi_X)$ with $s_\gamma (b)
= \langle p,\gamma (p)\rangle $ for any $p\in G_b$ and {\it viceversa},  $s
\in \Gamma (\xi_X)$ induces $\gamma_s \in \Gamma_{eq}(P,X), \gamma _s (p) = x$
where $s(\pi(p))=\langle p,x\rangle $. Pictorially,

\[
\diagram{
G&&\cr
\varr{}{}&&\cr
P&\harr{\gamma }{}&X\cr
\varr{\pi }{}&&|\cr
B&&P_X\cr
&&\udarr{s}{\pi_X}\cr
&&B\cr
}
\]

If P $\buildrel f \over \longrightarrow$ P is a gauge transformation of
$\xi$ {\it i.e.} an element of ${\cal G}$ $(\xi)$ and s $\in \Gamma (\xi_X)$
then the gauge transformation of s is defined to be

\begin{equation}
s':=s_{f^*(\gamma_s)}
\end{equation}

with $f^*(\gamma_s)$ given by the diagram

\[
\diagram{
&&X\cr
&f^*(\gamma _s)\nearrow &\uarr{}{\gamma _s}\cr
P&\harr{}{f}&P\cr
}
\]
then s'(b) $= \langle p,\gamma_s \circ f(p)\rangle  = \langle p,
\gamma_s(pg)\rangle  = \langle p,g^{-1} \cdot \gamma _s(p)\rangle $ with p$\in
G_b$ and g $\in G$ such that f(p) = pg.

For later applications, two important bundles canonically associated to a
p.f.b. $\xi$ are the following. Let {\bf g} be the Lie algebra of $G$.
$G\times G \buildrel {Ad} \over \longrightarrow G, (g,h)\mapsto Ad(g,h) = A_g
(h) = g\cdot h=ghg^{-1}$ and $G\times {\bf g} \buildrel {ad} \over
\longrightarrow {\bf g} ,(g,v)\mapsto ad(g,v)=g\cdot v := A_{g*e}(v) = dA_g|_e
(v) $ are the left {\it adjoint actions} of $G$ on itself and on {\bf g},
respectively. Associated with these actions one has the {\it bundle of Lie
groups of  $\xi$}, $\xi_G = (P_G = P \times_G G\equiv   F, B ,\pi _G, G)$ and
the {\it bundle of Lie algebras of $\xi$}, $\xi _{{\bf g}} =(P_G = P \times_G
{\bf g} \equiv  E, B , \pi_{{\bf g}} ,{\bf g})$  which is a real vector bundle
of rank r. It is easy to see that if $\Gamma (\xi _G) = \Gamma (F)$ is the
space of sections of $\xi _G $ with composition law $s \cdot s' (b) = \langle
p,gg'\rangle $ if $s(b) = \langle p,g\rangle $ and $s'(b)=\langle p,g'\rangle
,$ then $\Gamma_{eq} (P,G)  \buildrel \mu \over \longrightarrow \Gamma (\xi
_G)$ given by $\mu (\gamma) (b) := \langle p,  \gamma (p)\rangle $ with $ p
\in G_b$ is a group isomorphism , which provides the promised third equivalent
version of the gauge group: we have the isomorphism ${\cal G}(\xi) \buildrel
{\mu \circ \Sigma } \over \longrightarrow \Gamma (\xi _G)$. Moreover, $\Gamma
(\xi _{{\bf g}}) = \Gamma (E)$, the space of sections of $\xi _{{\bf g}}$, is
the {\it Lie algebra} of ${\cal G}(\xi)$ with exponential map $\Gamma ( \xi
_{{\bf g}}) \buildrel {Exp} \over \longrightarrow \Gamma (\xi _G)$ given by
$Exp (\sigma) (b) = \langle p, exp (v)\rangle $ if $\sigma(b) = \langle
p,v\rangle $ and $exp : {\bf g} \to G$ is the usual exponential function
associated with the Lie group $G$. We summarize this with the following
picture:

\[
\diagram{
G&&{\bf g}\cr
|&&|\cr
F&&E\cr
\duarr{\pi _G}{s}&\larr{Exp}{}&\udarr{\sigma }{\pi _{\bf g}}\cr
B&&B\cr
}
\]

Notice that locally ( or globally for trivial bundles) $P_\alpha \times _G
G\cong U_\alpha
\times G$ and $P_\alpha \times_G {\bf g} \cong U_\alpha \times {\bf g}$;
so ${\cal G}(\xi_\alpha) \cong \Gamma
((\xi_G)_\alpha) \cong C^\infty (U_\alpha,G):G$-valued smooth functions
on $U_\alpha$, and Lie $({\cal G}(\xi _\alpha ))\cong \Gamma ((\xi _{{\bf
g}})_\alpha )\cong C^\infty (U_\alpha ,{\bf g})$: ${\bf g}$-valued smooth
functions on $U_\alpha $.

The center of ${\cal G}$ is defined as the subset z of $\Gamma (\xi_G)$ such
that $s(b) = \langle p,g\rangle $ with $g \in  Z(G)$, the center of G; such
elements are well defined since $\langle p,g\rangle  = \langle ph,h^{-1} g
h\rangle  = \langle ph, g\rangle $. Clearly z is an invariant subgroup of
${\cal G}$ and one has the quotient group ${\cal G}$/z. For later use  the
subgroup $\bar z$ of z of constant $Z$-valued sections of $\xi_G$ is of
interest (in particular $\bar z = z$ for $SU(n))$, also $\bar {\cal G}:= {\cal
G}/\bar z$ is well defined.

For $ p=0,1,\ldots , s=dim$ $B$ define the infinite dimensional real vector
spaces $ A^p := \Gamma (\bigwedge^p T^* B \otimes E)$: differential
p-forms  on B with values in the Lie algebra of ${\cal G}(\xi)$ {\it i.e.}
if $\alpha \in A^p$ then $\alpha : \Gamma (T B)\times \ldots \times
\Gamma (T B) \to \Gamma (E) , (X_1,\ldots X_p) \mapsto  \alpha(X_1,\ldots ,X_p)
: B \to E; A^0 = \Gamma (\xi_{{\bf g}})$ and as we shall see in the next
section the affine space of connections on $\xi , {\cal C}(\xi)$ is modelled
on $A^1$ and therefore $A^1 \cong  T_\omega ({\cal C}, \omega_0 )$ for any
$\omega \in {\cal C}$ and arbitrary fixed $\omega _0 \in {\cal C}$ (base
point). After Sobolev completion the $A^p{'s}$ become complete inner product
linear spaces {\it i.e.} Hilbert spaces [8] with the inner products defined as
follows: i) let $G$ be a compact connected simply-connected
 Lie group e.g. $G = SU(2)$; then the Killing form on ${\bf g}$, ${\cal B}
:{\bf g} \times {\bf g} \to {\bf R}, {\cal B}(v,w) := - tr (ad (v) \circ ad(w))
$
with $ ad : {\bf g} \to End ({\bf g}), ad (v)(w) = [v,w]$ the adjoint
representation of ${\bf g}$ (ad is a Lie algebra homomorphism : $ad([v,w]) =
ad (v) \circ ad(w) - ad(w)\circ ad(v)$) is a positive-definite non-degenerate
symmetric bilinear form [11]; ${\cal B}$ induces the Killing-Cartan Riemannian
metric on $G$, $\langle v_g , w_g\rangle _g = {\cal B}(L_{g^{-
1}*g}(v_g),L_{g^{-1}*g}(w_g)), $ where $L_g$ is the left translation by $g$;
ii)
let B be a compact and orientable manifold; paracompactness and orientability
respectively guarantee the existence of a Riemannian metric and therefore of a
Hodge-*  operation  and of a volume form *1 on B; then the inner product on
$A^p$ is given by $\langle ,\rangle _p : A^p \times A^p \to {\bf R}, \langle
\alpha ,\beta \rangle _p := -\int_B tr (\alpha \land * \beta).$ Locally, $-
tr(\alpha \land *\beta )={1\over {p!}}g^{k_1l_1}\dots g^{k_pl_p}{\cal
B}(\alpha _{k_1\cdots k_p},\beta _{l_1\cdots l_p})\times \root \of
{det(g_{ij})}dx^1\land \cdots \land dx^s$ (repeated indices are summed from 1
to s and at each $b\in B$, $\alpha _{k_1\cdots k_p}(b)$, $\beta _{l_1\cdots
l_p}(b)\in {\bf g})$. For $\alpha \in A^p$, $\parallel \alpha \parallel
_p:=+\root \of {\langle \alpha ,\alpha \rangle _p}$. It can be shown that
these inner products are invariant under gauge transformations of $\xi $. If
$\alpha \in A^q$ its gauge transformed under $f\in {\cal G}(\xi )$ is defined
as follows: let $X_1,\ldots ,X_q\in \Gamma (TB)$, then $\alpha (X_1,\ldots
,X_q)\in \Gamma (\xi _{{\bf g}})$ and $\alpha (X_1,\ldots ,X_q)'(b)=\langle p,
A_{g^{-1}*e}\circ \gamma _{\alpha (X_1,\ldots ,X_q)}(p)\rangle $ with $p\in
G_b$. In particular $\parallel \alpha \parallel _q$ is gauge invariant for any
$\alpha \in A^q$.

Finally, let $\Gamma^k=\Gamma (\bigwedge ^kT^*P\otimes {\bf g})$ be the vector
space of differential k-forms on $P$ with values in ${\bf g}$. In applications
it is useful to consider the subspace $\bar \Gamma ^k=\bar \Gamma (\bigwedge
^kT^*P\otimes {\bf g})$ consisting of the k-forms $\phi $ satisfying the
conditions : i) $\phi _p(X_{1p},\ldots ,X_{kp})= 0\in {\bf g}$ if for some $j$,
$X_{jp}\in V_p=T_pG_b$, the {\it vertical space at p, i.e.} the $\phi 's$ are
horizontal; ii) for $X_1,\ldots ,X_k\in \Gamma (TP)$, $g\in G$ and $q=pg \in
P$, $\psi ^*_{gq}(\phi _q)(X_{1p},\ldots ,X_{kp})=\phi _q(\psi
_{g*p}(X_{1p}),\ldots ,\psi _{g*p}(X_{kp}))=A_{g^{-1}*e}\circ \phi
_p(X_{1p},\ldots ,X_{kp})$ {\it i.e.} under diffeomorphisms of $P$ induced by
the action of $G$ on $P$, the $\phi 's$ transform according to the adjoint
representation of $G$. ($\bar \Gamma ^k$ is usually called the space of $Ad-G$
invariant horizontal $k-$forms on $P$ with values in {\bf g}.) In particular
we will show in section 3 that the spaces $\bar \Gamma ^1$ and $A^1$ are
isomorphic.

\section{Space of connections}

There are four equivalent definitions of a connection on a p.f.b. or {\it
principal connection}. To be self-contained and for practical purposes, we
give here all the definitions leaving to the reader the details of the proof
of their equivalence. A {\it connection} on a $p.f.b.$ $\xi =(P,B,\pi
,G,{\bf U},\psi )$ is:

a) (Geometric definition) An assignment at each $p\in P$ of a vector
subspace $H_p$ (horizontal space) of $T_pP$ with the properties: {\it i})
$\pi _{*p}|_{H_p} : H_p\to T_{\pi(p)}B$ is a vector space isomorphism; {\it
ii})
${\psi}_{g*p}(H_p) = H_{{\psi}_g(p)}$; {\it iii}) for all $p\in $ $P$ there
exists open $U_p\subset P$ and a set of vector fields $
{V_{1q},...,V_{sq}}$ on $U_p$ such that for all $q\in U_p,
{V_{1q},...,V_{sq}}$ is a basis of $H_q$ {\it i.e.} the asignment $ p\mapsto
H_p$ is a smooth $s$ dimensional distribution on $P$. A consequence of this
definition is that $T_pP = H_p\bigoplus V_p$ and so for each $v_p\in T_pP$ the
decomposition $v_p = hor(v_p)\oplus ver(v_p)$  is unique.

b) (Algebraic definition) A differential 1-form on $P$ with values
in ${\bf g}$ {\it i.e.} an element ${\omega } \in {\Gamma }(T^*P\otimes {\bf
g})$  with the following properties:  i) ${{\psi }_g}^*_q(\omega _q) = A_{g^{-
1}*e}\circ \omega _p$ if $ q= pg$ ({\it i.e.} under the action of $G$ on $P$,
${\omega }$ transforms according to the adjoint representation of $G$);
 ii) $\omega _p(A^*_p) = A$ ,  for all $ p\in P$ and for all $ A\in {\bf g}$
with $ A^*_p\in V_p$ and $A^*\in  \Gamma (TG_b)$ the fundamental vector field
associated with $A$ and defined by $ A^* : G_b \to TG_b$, $A^*(p) =
(p,A^*_p)$, $A^*_p : C^{\infty }(G_b,{\bf R}) \to {\bf R}$, $A^*_p(f) =
{d\over dt} f(p$ $exp$ $tA)|_{t=0}$. The horizontal vector space at $p$ is then
defined as $H_p := ker(\omega _p)$  and $\omega _p(v_p) =\phi ^{-1}_p
(ver(v_p))$  with ${\bf g}\buildrel {\phi _p}\over \to V_p$ the canonical
vector space isomorphism given by $\phi _p(A) = A^*_p$. Notice that $\omega
_p(v_p) = 0$ iff $ v_p\in H_p$ ${\it i.e.}$ ${\omega }$ is a {\it vertical} 1-
form on $P$.

c) (Gauge theory definition)
A set $\{ A_\alpha \} _{\alpha \in J}$ of ${\bf g}$-valued differential 1-
forms on $\{ U_\alpha \} _{\alpha \in J}$ with ${\cal U}= \{ (U_\alpha , \phi
_\alpha )\} _{\alpha \in J}$ an atlas for ${\xi }$, related by  $ A_{\alpha b}
= A_{g^{-1}*e}\circ A_{\beta b} + L_{g^{-1}*g}\circ g_{\beta \alpha *b}$ for
each $b\in U_\alpha \cap U_\beta $ and where $ g= g_{\beta \alpha }(b)$ with
$g_{\beta \alpha }:U_\alpha \cap U_\beta \to G$ such that $\sigma _\alpha (b)
= \sigma _\beta (b)g_{\beta \alpha }(b)$. The $A_{\alpha }$'s are the {\it
gauge potentials} in physical applications and the $U_{\alpha }$'s are open
sets in space-time; usually $U_{\alpha } = B$  and ${\xi }$ is trivial. The
restriction ${\omega }_{\alpha }$ on $P_{\alpha }$ of the unique connection
${\omega }$ in ${\xi }$ determined by the $A_{\alpha }$'s satisfies $A_\alpha =
\sigma _\alpha ^*(\omega _\alpha )$ or $A_{\alpha b} = \omega _{\sigma
_\alpha(b)}\circ \sigma _{\alpha *b}$. The local relation between the
$A_{\alpha }$'s is called in physics  "gauge transformation"; however
according to the modern geometric terminology we shall reserve that name of
the elements of the global group ${\cal G}(\xi )$.

d) (Topological definition)
 A $G$-equivariant splitting ${\gamma }$ of the short exact sequence (s.e.s.)
of vector bundles over $P$, $0\to VP\buildrel i\over \to TP\buildrel {\tilde
\pi }\over \to \pi ^*TB\to 0$ where $VP=\coprod _{p\in P}V_p$ is the vertical
bundle of $P$, $i$ is the inclusion, $\pi ^*(TB)\subset P\times TB$  is the
pull-back bundle induced by the projection $P\buildrel \pi \over \to B$ {\it
i.e.}

\[
\diagram{
\pi ^*(TB)&\harr{p_2}{}&TB\cr
\varr{p_1}{}&&\varr{}{\pi _B}\cr
P&\harr{}{\pi }&B\cr
}
\]

and $\tilde{\pi}$ the bundle map induced by {\it i)} the map

\[
\diagram{
TP&\harr{\pi _*}{}&TB\cr
\varr{\pi _P}{}&&\varr{}{\pi _B}\cr
P&\harr{}{\pi }&B\cr
}
\]

and {\it ii)} the factorization property of the pull-back, leading to $\pi _* =
p_2\circ \tilde \pi $:

\[
\diagram{
TP&\harr{\tilde \pi }{}&\pi ^*(TB)&\harr{p_2}{}&TB\cr
&\pi _P\searrow &\varr{p_1}{}&&\varr{}{\pi _B}\cr
&&P&\harr{}{\pi }&B\cr
}
\]
$(\tilde \pi (p,v_p) = (p, \pi _*(p,v_p))$. For each $p\in P$, there exists
$(\pi ^*(TB))_p\buildrel {\gamma _p}\over \to T_pP$, a 1-1 linear
transformation of vector spaces satisfying $\tilde \pi _p\circ \gamma _p =
id_{(\pi ^*(TB))_p}$ and one has the linear isomorphism $V_p\oplus (\pi
^*(TB))_p\buildrel {\eta _p}\over \to T_pP,\eta _p(v_p\oplus w_b) = i_p(v_p) +
\gamma _p(w_p)$. Clearly $(\pi ^*(TB))_p$ is the horizontal space at $p\in P$
and $G$-equivariance of ${\gamma }$ means the condition {\it ii)} of
definition a). (Paracompactness of $P$ guarantees the existence of a
connection on ${\xi }$.)

        Let ${\cal C}$(or ${\cal C}(\xi ))$ denote the set of {\it all}
connections on ${\xi }$, its algebraic structure and topology will be
discussed later. Notice that ${\cal C}$ is a "natural" object associated with
${\xi }$. If ${\omega }_0$ is a fixed connection (base point) ${\cal C}$  is
denoted by ${\cal C}_0$ or $({\cal C},\omega _0)$. In the trivial bundle, the
{\it product connection} on $B\times G$ is canonically defined as follows: $TP
= TB \oplus TG =\coprod _{(b,g)\in B\times G} T_bB\oplus T_gG$ , so $H_{(b,g)}
:= T_bB$ for all $ g\in G$ $(V_{(b,g)} = T_gG)$; the connection form ${\omega
}_0$ is given by $\omega _{0_{(b,g)}} (v_b\oplus v_g) = \phi ^{-
1}_{(b,g)}(ver(v_b\oplus v_g)) =\phi ^{-1}_{(b,g)}(v_g):= L_{g^{-1}*g}(v_g) =
v_e\in {\bf g}$ {\it i.e.} $\phi^{-1}_{(b,g)} = L_{g^{-1}*g}$ for all $ b\in B$
and since $L_{g^{-1}*g} : T_gG\to {\bf g}$ is canonical we can identify
$\omega _{0(b,g)}(v_b\oplus v_g) \equiv v_g$. For the trivial bundle
$B\buildrel \sigma \over \to B\times G, \sigma (b)= (b,e)$ is a global
canonical section and if $\omega$ is an arbitrary connection, $A_b =
\omega_{(b,e)}\circ \sigma _{*b}$ does not vanish in general; however for the
product connection the gauge potential vanishes: in fact, $A_b^0(v_b)=\omega
_{0(b,e)}\circ \sigma _{*b}(v_b)\equiv ver(\sigma _{*b}(v_b))$ according to
the previous identification and $\sigma _{*b}(v_b)$ is horizontal since
$T_bB\buildrel {\sigma _{*b}}\over \to T_bB\oplus {\bf g}$, $\sigma _{*b}(v_b)
= v_b\oplus 0$, so $A^0 = \sigma ^*(\omega _0) = 0$.

        A connection on a p.f.b. immediately leads to the concept of {\it
horizontal lifting} of vector fields: let $X$ be a vector field on $B$,
then $X_b \in T_bB$ and we define the horizontal lifting $\tilde X\in \Gamma
(TP)$ by $\tilde X_p =(\pi _*|_{H_p})^{-1}(X_b)\in H_p$.

         Let ${\omega }\in {\cal C}$ and  $f\in {\cal G}(\xi )$, then it is
easy to prove that the pull-back $f^*(\omega )$, the {\it gauge transformed
connection}, is also an element of ${\cal C}$; a little more work shows that
$f^*(\omega )$ is given by

\begin{equation}
f^*(\omega _{f(p)}) = (L_{\gamma (p)^{-1}}\circ \gamma)_{*p} + A_{\gamma (p)^{-
1}*e}\circ  \omega _p
\end{equation}
which belongs to $T^*_pP\otimes {\bf g}$   and where $\gamma = \Sigma (f)\in
{\Gamma }_{eq}(P,G)$. The right hand side of (2) is given by the sum of the two
diagonals in the following diagrams:

\[
\diagram{
&&{\bf g}\cr
&\nearrow &\uarr{}{L_{\gamma (p)^{-1}*\gamma (p)}}\cr
T_pP&\harr{}{\gamma _{*p}}&T_{\gamma (p)}G\cr
}
\]
\[
\diagram{
&&{\bf g}\cr
&\nearrow &\uarr{}{A_{\gamma (p)^{-1}*e}}\cr
T_pP&\harr{}{\omega _p}&{\bf g}\cr
}
\]

In physical language, the formulas (1) and (2) summarize the gauge
transformations of matter and gauge fields. Locally, for matrix Lie groups
(2) is given by the well known formula

\begin{equation}
\omega ' = g^{-1}\omega g  + g^{-1}dg
\end{equation}
where $g\in C^{\infty }(U_{\alpha },G)$.

        The {\it curvature} of the connection ${\omega }$ is the differential
2-form on $P$ with values in ${\bf g}$ given by $\Omega = {\cal D}\omega :=
(d\omega )^{hor}\in \Gamma (\Lambda ^2T^*P\otimes {\bf g})$ with $(d\omega )
^{hor}(X,Y)=d\omega (horX, horY)$; clearly ${\Omega}$ is horizontal {\it i.e.}
$\Omega _p(X_p,Y_p) = 0 $ if $ X_p$ and/or $Y_p\in V_p$ and one can prove that
$\Omega = d\omega +{1\over 2}[\omega ,\omega ]_{\wedge }$ where  ${1\over
2}[\omega ,\omega]_{\wedge}(X,Y)$ denotes ${1\over 2} \omega^a \wedge \omega^b
(X,Y) [e_a,e_b]$ $= {1\over 2}(\omega^a(X) \omega^b(Y)$ $- \omega^a(Y)
\omega^b(X))$ $[e_a,e_b]$ in a basis $\{e_a\}_{a=1}^{dim G}$ of {\bf g}, for a
matrix Lie group ${1\over 2}[\omega,\omega]_{\wedge}(X,Y)$ $= [\omega (X),
\omega (Y)]$. A connection ${\omega }$ for which ${\cal D}\omega =0$ is called
{\it flat}. Clearly, the product connection on the trivial bundle is flat.
Under a gauge transformation $f\in {\cal G}(\xi )$,

\begin{equation}
f^*(\Omega _{f(p)}) = A_{\gamma (p)^{-1}*e}\circ \Omega _p \in
\Lambda ^2T^*_pP\otimes {\bf g}
\end{equation}
which locally and for matrix groups reduces to the formula

\begin{equation}
\Omega ' = g^{-1}\Omega g
\end{equation}
with $g\in C^{\infty }(U_{\alpha },G)$.

Let ${\omega }$ be a connection on $P$, ${\Omega }$ its curvature, $X,Y\in
\Gamma (TB)$ and $\tilde X$, $\tilde Y$ their horizontal liftings. One can
prove that ${\Omega }(\tilde X,\tilde Y)\in {\Gamma }_{eq}(P,{\bf g})$, then
$\tilde \Omega (X,Y) := s_{\Omega (\tilde X,\tilde Y)}$   defines an element
$\tilde \Omega \in A^2 = \Gamma (\Lambda ^2T^*B\otimes E)$. The {\it Yang-
Mills action} is the function $YM : {\cal C}\to {\bf R}$, $\omega \mapsto
YM(\omega) := $\\ $(\parallel \tilde \Omega \parallel _2)^2$ whose extrema
(critical points in the sense of Morse theory) give the solutions of the
classical equations of motion

\begin{equation}
 {\cal D}^{2*} \tilde \Omega = 0
\end{equation}
(see below the definition of the covariant derivative and divergence). By
gauge invariance of the inner product $\langle ,\rangle _2$, {\it YM} is indeed
a function on the quotient space  ${\cal C}/{\cal G}$  (more precisely ${\cal
C}'/{\bar {\cal G}}$, see below), whose topology is non-trivial due to the
non-trivial homotopy of the gauge group ${\cal G}$. Finally, $\Omega$
satisfies the Bianchi identity ${\cal D}\Omega = (d\Omega )^{hor}=0$; from
the physical point of view one can say that the "half" of the classical
equations of motion for the gauge fields has purely geometric origin (Bianchi
identity), while the "other half" or "dynamical equations" (6) are a
consequence of the somewhat arbitrary definition of $YM$.

Let ${\xi }_V : V - P_V\buildrel {\pi _V}\over \to B$ be a real vector bundle
associated with $ {\xi } : G\to P\buildrel \pi \over \to B$, ${\omega }$ a
connection on ${\xi }$ and $s$ a section of ${\xi }_V$. As we saw in section
2, $s$ induces $\gamma _s\in \Gamma _{eq}(P,V)$ and if $X$ is a vector field
on $B$ then one can prove that $\tilde X(\gamma _s)$ is also equivariant {\it
i.e.} $\tilde X(\gamma _s)\in \Gamma _{eq}(P,V)$, which induces the {\it
covariant derivative} of s with respect to ${\omega }$ in the direction $X$

\begin{equation}
\nabla ^{\omega }_V(X,s) := s_{\tilde X(\gamma _s)}
\end{equation}

(For $f\in \Gamma _{eq}(P,V)$, $Y\in \Gamma (TP)$ and $\omega \in {\cal C}$,
the covariant derivative of $f$ with respect to $\omega $ in the direction $Y$
is defined as $Df(Y):=df(horY)$, clearly if $Y$ is horizontal $Df(Y)=Y(f)$;
for $G\times V\buildrel \mu \over \to V$ a linear action and $p\in P$,
$(Df)_p(Y_p)=f_{*p}(Y_p)+\tilde \mu _{*e}(\omega _p(Y_p))(f(p))\in V$ with
$\tilde \mu :G\to GL(V)$ given by $\tilde \mu (g):=\mu _g$.)

Diagramatically

\newpage

\[
\diagram{
&&G&&\cr
&&\varr{}{}&&\cr
TP&\rlarr{\pi _p}{\tilde X}&P&\harr{}{\omega }&T^*P\otimes {\bf g}\cr
&&\varr{\pi }{}&&\cr
TB&\rlarr{\pi _B}{X}&B&&\cr
}
\]
\[
\diagram{
P&\rrarr{\gamma _s}{\tilde X(\gamma _s)}&V\cr
&&|\cr
&&P_V\cr
&&\udarr{\nabla _V^\omega (X,s)}{\pi _V}\cr
&&B\cr
}
\]

For brevity one omits $V$ and ${\omega }$ in the symbol of the covariant
derivative and writes $\nabla ^{\omega }_V(X,s) = \nabla _Xs$. It is easy to
verify that the operator $\nabla ^{\omega }: \Gamma (TB)\times \Gamma (P_V)
\to \Gamma (P_V), \nabla ^{\omega }(X,s) := \nabla _X s$ is a {\it linear
connection} in ${\xi }_V$ {\it i.e.} $\nabla _{X+X'}s = \nabla _X s + \nabla
_{X'}s$, $\nabla _{fX}s = f\nabla _Xs$, $\nabla _X(s + s') = \nabla _Xs +
\nabla _Xs'$ and $\nabla _X(fs) = X(f) s + f\nabla _X s$ for any $X,X' \in
\Gamma (TB), s,s'\in \Gamma (P_V)$ and $ f\in C^{\infty }(B,{\bf R})$. In
other words, $\nabla ^{\omega }$ is $C^{\infty }(B,{{\bf R}})$-linear with
respect to $X$  but satisfies the Leibnitz rule with respect to s. The {\it
curvature} of the linear connection $\nabla ^{\omega }$ is the operator ${\cal
R}^{\omega } : \Gamma (TB)\times \Gamma (TB)\to End(\Gamma (P_V)),
(X,Y)\mapsto {\cal R}^{\omega }(X,Y) : \Gamma (P_V)\to \Gamma (P_V), s\mapsto
{\cal R}^{\omega }(X,Y)(s):= ([\nabla _X,\nabla _Y] - \nabla _{[X,Y]})(s)$
with $[\nabla _X,\nabla _Y] = \nabla _X\circ \nabla _Y - \nabla _Y\circ \nabla
_X$. ${\cal R}^\omega $ is $C^\infty (B,{\bf R})$-linear with respect to $X$,
$Y$ and $s$ {\it i.e.} ${\cal R}^\omega (fX,Y)(s)={\cal R}^\omega
(X,fY)(s)={\cal R}^\omega (X,Y)(fs)=f{\cal R}^\omega (X,Y)(s)$.

In terms of the previously defined covariant derivative one defines the
(linear) {\it covariant derivative operator} with respect to ${\omega }$ in the
associated bundle ${\xi }_V$, $ d_\omega : \Gamma (P_V)\to \Gamma (T^*B\otimes
P_V)$, $s\mapsto d_{\omega }s : \Gamma (TB)\to \Gamma (P_V), d_{\omega
}s(X):= \nabla _X s$. One easily verifies that $d_\omega (fs)=(df)s+fd_\omega
s$. (Again the full notation should be $d_{\omega}^V)$. From the physical
point of view the object $d_\omega s$ establishes the {\it interaction}
between the matter field $s$ and the gauge field ${\omega }$. In the following
and with the purpose of practical applications we shall restrict the
discussion to the case $ V = {\bf g}$ (one could of course mantain the
discussion at a general level for an arbitrary associated vector bundle ${\xi
}_V$).

        In the same way as the De Rham exterior derivative generalizes the
concept of differential of a function, the concept of {\it covariant exterior
differentiation} generalizes the covariant derivative of sections of ${\xi
}_{{\bf g}}$. The obvious spaces which replace the spaces $\Omega ^p(B)$ of
differential p-forms on B are the spaces $A^p$ defined in section 2; then one
defines ${\cal D}^p : A^p\to A^{p+1}$ with ${\cal D}^0 = d_\omega $ and for
\hskip5pt
$p=1,...,s, {\cal D}^p(\alpha )(X_1,...,X_{p+1}) =
\sum _{i=1}^{p+1}(-1)^{i+1}\nabla
_{X_i}(\alpha (X_1,...,\hat X_i,$ \\ $...,X_{p+1})) + \sum _{1\leq
i<j\leq p+1}(-1)^{i+j}\alpha
([X_i,X_j],X_1,...,\hat X_i,...,\hat
X_j,...X_{p+1})$ . Comparing with the objects defined in
Appendix A which lead to the Chevalley-Eilenberg (C-E) cohomology of a Lie
algebra [12], one is tempted to identify ${\cal C}^p = A^p$, ${\bf g} =
\Gamma (TB)$, $V = \Gamma (E)$  and $\delta ^p = {\cal D}^p$ ; however
contrary to the case in De Rham theory the sequence $ 0\to A^0
\buildrel {d_\omega }\over \to A^1\buildrel {{\cal D}^1}\over \to
A^2\buildrel{{\cal D}^2}\over\to \ldots\buildrel {{\cal D}^{s-1}}\over \to
A^s\to 0$ is {\it not} in general a complex since $\{ {\cal D}^p\} _{p=0}^{s-
1}$   is not a coboundary operator {\it i.e.}  ${\cal D}^{p+1}\circ {\cal
D}^p$ does not vanish in general. The necessary and sufficient condition to
have a coboundary and therefore a C-E cochain complex is that the operator
$\delta ^{\omega } : \Gamma (TB)\to End(\Gamma (E)), X\mapsto \delta ^{\omega
}(X) : \Gamma (E)\to \Gamma (E), s\mapsto \delta ^{\omega }(X)(s):= \nabla
_Xs$ be a Lie algebra representation, but from the definition of ${\cal R
}^{\omega }$, $\delta ^{\omega }([X,Y]) = [\delta ^{\omega }(X),\delta
^{\omega }(Y)] -{\cal R}^{\omega }(X,Y)$ and thus we have the result that for
each {\it flat} connection ${\omega }$ (if any) on a p.f.b. for which ${\cal
R}^{\omega } = 0 $ one has a C-E complex {\it i.e.} the set of operators $\{
{\cal D}^p\} _{p=0}^{s-1}$ does indeed satisfy ${\cal D}^{p+1}\circ {\cal D}^p
= 0$ and therefore a C-E cohomology $H^*_{CE}(\Gamma (TB)$, $\delta ^{\omega
}$, $Lie({\cal G}(\xi );{{\bf R}})$. One says that the curvature of the
connection is an {\it obstruction} to the existence of the cohomology of the
Lie algebra $\Gamma (TB)$ with coefficients in $\Gamma (E)$.

        Independiently of the flatness or not of the connection ${\omega }$,
one can prove that each $ {\cal D}^p$   has an {\it adjoint} operator with
respect to the inner product $\langle ,\rangle _p$  in $A^p, {\cal D}^{p*} :
A^{p+1}\to A^p$, called {\it exterior covariant divergence} for $p\geq 1$ and
{\it covariant divergence} for $p=0$, such that for any $\alpha \in A^p$ and
$\beta \in A^{p-1}, \langle \alpha ,{\cal D}^{p-1}\beta \rangle _p=\langle
{\cal D}^{p*}\alpha ,\beta \rangle _{p-1}$. (The construction of the adjoints
runs according to the following general procedure: Let $V$ and $W$ be vector
spaces with inner products $\langle ,\rangle _V$ and $\langle ,\rangle _W$
which induce isomorphisms $V \buildrel \mu _V \over \longrightarrow V^*$ and
$W \buildrel \mu _W \over \longrightarrow W^*$ given by $\mu _V(v)(v')=\langle
v,v'\rangle _V$ and $\mu _W(w)(w')=\langle w,w'\rangle _W$. This is the case
e.g. of Hilbert spaces in the infinite dimensional case or for arbitrary
finite dimensional inner product spaces. If $V \buildrel f \over
\longrightarrow W$ is a linear transformation (as $A^p \buildrel {\cal D}^p
\over \longrightarrow A^{p+1}$ is),  $W^* \buildrel \bar f \over
\longrightarrow V^*$ given by $\bar f( \gamma ) = \gamma \circ f$ closes the
diagram

\[
\diagram{
V&\harr{\mu _V}{}&V^*\cr
\duarr{f}{f^*}&&\uarr{}{\bar f}\cr
W&\harr{}{\mu _W}&W^*\cr
}
\]
and defines the composition $f^* = \mu _V ^{-1} \circ \bar f \circ \mu _W$
which satisfies \\ $\langle f(v),w\rangle _W=\langle v,f^*(w)\rangle _V$.) One
has the sequence \[ 0 \longleftarrow A^0 \buildrel d_\omega ^* \over
\longleftarrow A^1
   \buildrel {\cal D}^{1*} \over \longleftarrow A^2\buildrel {\cal D}^{2*}\over
   \longleftarrow \cdots \buildrel {\cal D}^{s-2*}\over \longleftarrow A^{s-1}
   \buildrel {\cal D}^{s-1*} \over \longleftarrow A^s \longleftarrow 0
\]
which again in general is not a chain complex; clearly $\{ {\cal D} ^{p*} \}
_{p=0}^{s-1}$ is a {\it boundary i.e.} ${\cal D} ^{p-1*} \circ {\cal D} ^{p*}
= 0$ iff $\{ {\cal D} ^{p} \} _{p=0}^{s-1}$ is a coboundary. (${\cal D} ^{p*}$
generalizes the codifferential $d^*$ in De Rham theory.) At each $p$ one
defines the generalized {\it Laplace-Beltrami} operator $\Delta _p := {\cal D}
^{p*} \circ {\cal D} ^{p} +
 {\cal D} ^{p-1} \circ {\cal D} ^{p-1*}$
(pictorially

\[
\cdots A^{p-1}
\rlarr{{\cal D}^{p-1}}{{\cal D}^{p-1^*}} A^p
\rlarr{{\cal D}^p}{{\cal D}^{p*}} A^{p+1}\cdots
=\cdots A^p\cdots ),
\]
in particular
$\Delta _0 = d_\omega ^* \circ d_\omega : A^0 \longrightarrow A^0 $ with
inverse (if it exists) the Green function
$G_\omega := \Delta _0^{-1}$.

Let $\bar \Gamma ^1$ be the space of horizontal $Ad-G$ invariant 1-forms on
$P$ with values in ${\bf g}$, and ${\cal C}$ the set of connections on $P$. It
is easy to show that $\bar \Gamma ^1  \times {\cal C}\buildrel \hat + \over
\longrightarrow  {\cal C}, \alpha \hat + \omega := \alpha + \omega $, where
the sum in the right hand side is the one in $\Gamma (T^*P\otimes {\bf g})$,
is a free and transitive action of $\bar \Gamma ^1$ on ${\cal C}$; then by
definition $(\bar \Gamma ^1,{\cal C},\hat +)$ is an {\it affine space} and if
$\omega _0$ is a distinguished connection (base point) in ${\cal C}$, $\mu _0
: \bar \Gamma ^1 \longrightarrow {\cal C}_0, \mu _0 (\tau ) := \tau
 \hat + \omega _0$ is a bijection with inverse
$\mu _0^{-1}( \omega ) = \omega - \omega _0$. Therefore
$({\cal C}_0, \tilde + ; {\bf R}, \tilde \cdot)$ with
$\omega \tilde + \omega ' := \mu _0^{-1}( \omega ) + \mu _0^{-1}
 (\omega ') \hat + \omega _0$ and
$\lambda \tilde \cdot \omega := \lambda \mu _0^{-1}( \omega ) \hat +
 \omega _0$ is a real infinite dimensional {\it vector space}. (In
particular one has the "straight line" of connections through
$\omega _1$ and
$\omega _0$ for arbitrary
$\omega _1$ in
${\cal C}_0$ given by
$\omega (t) = \omega _0 + t(\omega _1 - \omega _0) = (1-t)\omega _0 +
 t\omega _1$.) Notice that for
the product bundle
$B\times G$ the affine space of connections can be {\it identified} with a
vector space since the product connection is canonical; this is not the case
however in an arbitrary principal bundle.

Giving ${\cal C}_0$ the limit topology of
${\bf R}^\infty $ makes it a topological space of the homotopy type of a point
{\it i.e.} contractible and therefore with zero homotopy groups. The choice of
a vector space basis provides
${\cal C}_0$ with a global chart and makes it an infinite dimensional
differentiable manifold with tangent space
$T_\omega {\cal C}_0$ at each
$\omega \in {\cal C}_0$ isomorphic to
$\bar \Gamma ^1$, the differentiable structure is however independent of the
choice of basis and of $\omega_0$. The function $\rho _0 :{\cal C}_0
\longrightarrow A^1, \omega \mapsto \rho _0 (\omega ) :
 \Gamma (TB) \longrightarrow \Gamma (E), \rho _0 (\omega )(X) := s_{\omega
(\tilde X_0)}$ where
$\tilde X_0$ is the horizontal lifting of $X$ by
$\omega _0$ and
$s_{\omega (\tilde X_0)}$ the element in
$Lie({\cal G}(\xi ))$ corresponding to the equivariant
$\omega (\tilde X_0)$ in
$\Gamma _{eq}(P, {\bf g})$, turns out to be a bijection, and therefore one
has the composition

\[
\diagram{
&&\bar \Gamma ^1\cr
&\mu _0 \swarrow &\varr{}{\rho _0\circ \mu _0}\cr
{\cal C}_0&\harr{}{\rho _0}&A^1\cr
}
\]
which establishes a 1-1 and onto linear transformation between the vector
spaces
$\bar \Gamma ^1(T^*P\otimes {\bf g})$ and
$\Gamma (T^*B\otimes E)$; the isomorphism however is not canonical (except for
the trivial bundle
$B\times G$) since
$\rho _0 \circ \mu _0$ depends on
$\omega _0$.

One can prove that the infinite dimensional universal principal bundle
$\bar {\cal G} \longrightarrow {\cal C}' \longrightarrow {\cal C}'
 /\bar {\cal G} $ where
${\cal C}'$ is the subspace of
${\cal C}$ consisting of {\it irreducible} connections
($\omega $ is irreducible if any two points
$p$ and $p'$ in $P$
can be joined by a horizontal curve) is {\it not trivial} since
$\bar {\cal G}$ has at least one non-zero homotopy group [13] (see also [14]);
this leads to the {\it imposibility of a global gauge fixing, i.e.} of a
continuous choice of a representative for each gauge equivalence class of
connections in
${\cal C}'/\bar {\cal G}$. (The concept of irreducibility is closely related
to that of parallel transport. For any two points
$b$ and $b'$ in $B$
and a smooth curve
$\gamma $ joining them with
$\gamma (0) = b$ and
$\gamma (1) = b'$ it can be proved that there exists a unique horizontal
lifting
$\bar \gamma _p$ of
$\gamma $ in $P$ (all tangent vectors to
$\bar \gamma _p$ are horizontal and
$\pi \circ \bar \gamma _p = \gamma $) passing through any
$p \in G_b$. Then there exists the diffeomorphism
$\tau _\gamma : G_b \longrightarrow G_{b'}, \tau _\gamma (p) = \tau _\gamma
 (\bar \gamma _p (0)) := \bar \gamma _p (1)$ which is called the {\it
parallel transport of $G_b$ in $P$ through $\gamma $ in B}. From the physical
point of
view one can imagine that when a particle is classically transported through
the path $\gamma $ in space-time $B$ in the presence of the fields $\{
A_\alpha \} _{\alpha \in J}$ from the point $b$ to the point $b'$ its
"internal state" changes from $\bar \gamma _p (0) = p$ (initial state) to
$\bar \gamma _p (1)$ (final state). For a closed curve $\gamma $, there exists
and is unique $g \in G$ such that $\bar \gamma _p (1) = pg$ , and for the loop
space of $B$ at $b$, $\Omega (B,b)$ the corresponding set of parallel
transports $G_b \longrightarrow G_b$ form a group $\Phi _b$ , the {\it
holonomy group of the connection $\omega $ at the point $b$}. It can be shown
[15],[16] that for each $p \in G_b$ there exists a group isomorphism $J_p :
\Phi
_b \longrightarrow \phi _p$ where $\phi _p$ is the Lie group (closed subgroup
of $G$) given by $\phi _p = \{ g\in G | \exists \gamma \in \Omega (B,b) | \tau
 _\gamma (p) = pg \} $, {\it the holonomy group of the connection
$\omega $ with reference point $p$}. Also, if
$p$ and $p'$ can be joined with an horizontal curve then
$\phi _p = \phi _{p'}$ and if all points of $P$ can be joined with
horizontal curves to a given fixed point $p_0 \in P$ , then
$\phi _p = G$ for all $ p \in P$.)

Let $\omega \in {\cal C}_0$ and
$\eta \in A^0 = \Gamma (E) = Lie({\cal G}(\xi ))$, then
$d_\omega \eta \in A^1$ and
$(\rho _0 \circ \mu _0)^{-1} (d_\omega \eta ) \in \bar \Gamma ^1$ . So
$\omega '= \omega + d_\omega ^0 \eta $ is in
${\cal C}^0$ with
$d_\omega ^0 := (\rho _0 \circ \mu _0)^{-1} \circ d_\omega $;
$\omega '$ is the {\it Lie algebra} or {\it infinitesimal
transformation} of
$\omega $ generated by
$\eta $ . Notice that the "effective" covariant derivative operator
$d_\omega ^0$ depends on
$\omega _0$ , the distinguished element in
${\cal C}$; only for trivial bundles or locally for arbitrary bundles
$d_\omega $ and
$d_\omega ^0$ can be identified and one has the usual formula
$\omega ' = \omega + d_\omega \eta $ which can be formally obtained from
(3) as the
${\cal O}(t)$ term in the expansion of
$g = exp(t\eta )$ at $t=1$ considering
${\cal G}(\xi )$ as a group of matrices and identifying
$d\eta + [\omega ,\eta ]$ with
$d_\omega \eta $. Similarly as in finite dimensional Lie groups, for each
$\eta \in Lie({\cal G}(\xi ))$ one has the {\it one parameter group} of
infinitesimal transformations of
$\omega , \omega (t) = \omega + t d_\omega ^0 \eta $.

\section{BRST cohomology}

As we mentioned in the previous section the group of the gauge transformations
on a p.f.b., ${\cal G}(\xi )$ has a right action on the space of connections
${\cal C}(\xi )$, namely ${\cal C}\times {\cal G}\harr{\Phi }{}{\cal C}$,
$(\omega ,f)\mapsto \Phi (\omega ,f):=f^*(\omega )$ {\it i.e.} the action is
given by the gauge transformations of the connections.

As first observed in ref. [19] this action leads to the Chevalley-Eilenberg
cohomology of the Lie algebra of ${\cal G}(\xi )$ with coefficients in the
real valued functions (0-forms) on ${\cal C}(\xi )$, which was identified with
the Becchi-Rouet-Stora-Tyutin (BRST) cohomology [20] previously found in the
context of the quantum theory of gauge fields as a consequence of the global
symmetry naturally appearing at the end of the quantization procedure
{\it via} the method of path integration [1], the generator of the symmetry
precisely being the coboundary operator of the BRST complex. (Immediately
after the discovery in [20], an anti- BRST symmetry was found within the same
context by Curci and Ferrari [25], and it was until 1982 when Alvarez-Gaum\' e
and Baulieu [26] made the deep assertion that the full BRST symmetry is the
gauge symmetry in the perturbative quantum theory; moreover they proved that
the gauge invariance of the physical S-matrix elements is equivalent  to the
full BRST invariance (though not gauge invariance) of the "quantum
Lagrangian"; this point of view was extensively reviewed by Baulieu in [27].)

There are at least two reasons why the approach to the BRST symmetry from the
geometrical point of view is interesting. First, the BRST cohomology turns
out to be a property of the chosen principal bundle $\xi $ {\it i.e.} of the
"space" where the gauge fields live; in this sense it can be considered as a
purely geometrical property, independent of any lagrangian theory that one can
set up on the base space of the bundle (in particular of the YM lagrangian).
Second, the BRST cohomology groups contain information about the quantum
theory of any gauge theory that one can place on the principal bundle, in fact
the gauge {\it anomalies} are contained in the cohomology groups with ghost
number larger than or equal to one, as it is discussed in references [19] and
[21]. (Local ghost and anti-ghost fields, scalar particles with Fermi
statistics, naturally appear in the path integral quantization when one
expresses the Faddeev-Popov determinant in terms of the algebraic generators
of an infinite dimensional complex Grassmann algebra with involution. In the
geometrical description {\it the} (global) ghost field is the Maurer-Cartan
form of the gauge group.) This fact again suggests a deep connection between
the {\it topology} of fiber fundles and the {\it quantum theory} of gauge
fields (connections) as it was recently emphasized, though in an apparently
different context, by Atiyah [22] and Witten [23]. (Incidentaly in the recent
literature([21],[24]) there are efforts towards the  explicit calculation of
the BRST cohomology groups for several different concrete situations using the
technique of spectral sequences.)

Using the proposition of Appendix C one  makes the identifications $G={\cal
G}(\xi )$ and $M={\cal C}(\xi )$; then for each $p\in \{ 0,1,2,\ldots \} $ one
has the representation of the Lie algebra of the gauge group on the
differential p-forms over the space of connections given by $A^0=Lie({\cal
G}(\xi ))\buildrel {\phi _p}\over \to End(\Omega ^p({\cal C}(\xi )))$, $\phi
_p(s)={\cal L}_{s^*}$ where $s^*$ is the fundamental vector field on
${\cal C}(\xi )$ associated with $s$ and ${\cal L}_{s^*}$ is the Lie
derivative with respect to $s^*$, so for any real valued $f\in \Omega ^0({\cal
C}(\xi ))$, $s^*_\omega (f)= {d\over dt}f\circ \Phi (\omega ,(\mu \circ \Sigma
)^{-1}(Exp$ $ts))|_{t=0}={d\over dt}f(\omega +t d^0_\omega s)|_{t=0}$ where
the last expression for $s^*_\omega (f)$ depends on the base point $\omega _0$
in ${\cal C}$ (as mentioned in section 3 for trivial bundles $\omega _0$ is
canonical and $d^0_\omega $ can be identified with $d_\omega $). Defining the
spaces ${\cal C}^\nu _p(\xi )$ of alternating continuous (see below) functions
$A^0\times \ldots \times A^0$ ($\nu $ times)$\harr{\alpha }{}\Omega ^p({\cal
C}(\xi ))$ for $\nu =1,2,3,\ldots $ and ${\cal C}^0_p(\xi )=\Omega ^p({\cal
C}(\xi ))$ for $\nu =0$, with ${\cal C}(\xi )$ and $A^0$ respectively given
suitable Sobolev $k$- and ($k+1$)-norm completions with integer $k\ge [{dim
B\over 2}]+1$ (these completions guarantee that the action ${\cal C}\times
{\cal G}\to {\cal C}$ extends to a smooth action ${\cal C}_k\times {\cal
G}_{k+1}\to {\cal C}_k$ [8]; see also reference [28] for the case $k=3$,
$G=SU(2)$ and $B$ a compact 4-manifold), one has the double complex

\[
\diagram{
{\cal C}^0_0(\xi )&\harr{d^0_0}{}&{\cal C}^0_1(\xi )&\harr{d^0_1}{}&{\cal
C}^0_2(\xi )&\harr{d^0_2}{}&\cdots \cr
\varr{\delta ^0_0}{}&&\varr{\delta ^0_1}{}&&\varr{\delta ^0_2}{}&\cr
{\cal C}^1_0(\xi )&\harr{d^1_0}{}&{\cal C}^1_1(\xi )&\harr{d^1_1}{}&{\cal
C}^1_2(\xi )&\harr{d^1_2}{}&\cdots \cr
\varr{\delta ^1_0}{}&&\varr{\delta ^1_1}{}&&\varr{\delta ^1_2}{}&\cr
\vdots &&\vdots &&\vdots \cr
}
\]
with differentials  \\
$\delta _p^\nu : {\cal C}_p^\nu (\xi )\to {\cal C}_p^{\nu + 1}(\xi )$,
$\delta _p^\nu (\alpha )(s_0, \ldots ,s_\nu ) = $
$\Sigma _{i=0}^\nu (-1)^i {\cal L}_{s^*} (\alpha (s_0,\ldots
,\hat s_i,\ldots ,s_\nu ))$
$+ \Sigma _{0\leq i<j\leq \nu
}(-1)^{i+j} \alpha ([s_i,s_j],s_0,\ldots ,\hat
s_i,\ldots ,\hat s_j,\ldots ,s_\nu )$
and
$d_p^\nu : {\cal C}_p^\nu (\xi )\to {\cal C}_{p+1}^\nu (\xi )$,
$d_p^\nu (\alpha )(s_1, \ldots , s_\nu ) = d(\alpha (s_1,
\ldots , s_\nu ))$
where $d$ is the De Rham operator on the infinite dimensional manifold ${\cal
C}$ (the double complex is like that in appendix C, except that now it extends
to infinity also in the horizontal direction). The continuity condition for
$\alpha $ is given as follows: if $\alpha \in {\cal C}_p^\nu (\xi )$ then the
map $\alpha _{\omega ,\xi _1,...,\xi _p}:A^0\times \cdots \times A^0(\nu
\hbox{ }times)\to {\bf R}$, $(s_1,...,s_\nu )\mapsto \alpha _{\omega ,\xi
_1,...,\xi _p}(s_1,...,s_\nu ):=\alpha (s_1,...,s_\nu )(\xi _1,...,\xi
_p)(\omega )$ is continuous for all $\omega \in {\cal C}_k$ and $\xi
_1,...,\xi _p\in Vect({\cal C}_k)$. The usual BRST complex [19] is the
Chevalley-Eilenberg complex consisting of the first column in the previous
"lattice" which defines the {\it BRST cohomology of the principal bundle} $\xi
$ as $H_{BRST}^*(\xi )$ \\ $=\oplus_{\nu =0}^\infty H_{BRST}^\nu (\xi )=\oplus
_{\nu =0}^\infty {ker\delta _0^\nu \over im\delta _0^{\nu -1}}$ $(\delta _0^{-
1}=0)$, $\nu $ is the ghost number and the coboundary $\{ \delta _0^\nu \}
_{\nu =0}^\infty$ is identified with the BRST nilpotent operator appearing in
the quantum theory. The columns corresponding to $p=1,2,\ldots $ and the
corresponding differentials have been defined here in a formal way and we do
not have yet a physical interpretation (if any) of them. Following Appendix B
one has an associated total complex $(K,D)$ with $K^n(\xi )=\oplus _{\nu
+p=n}{\cal C}_p^\nu (\xi )$ and $D^n=\oplus _{\nu +p=n}(\delta _p^\nu \oplus
(-1)^\nu d^\nu _p)$ and therefore a total cohomology $H^*(K,D)$ which we might
call the {\it total BRST cohomology of a p.f.b. $\xi $} and denote by ${\cal
H}^*_{BRST}(\xi ).$ We believe that the possible implications and
interpretations of this definition deserves further research.

In reference [19] a geometric interpretation of the above Lie algebra
cohomology in terms of the vertical part of the De Rham exterior derivative on
the space of irreducible connections is given; however we believe that an
interpretation in terms of the topological (Eilenberg-Steenrod) cohomologies
of the relevant spaces of the bundle (Lie group, total space and base space)
should be more conclusive towards establishing a relationship between quantum
mechanics and topology [29].

\section*{Acknowledgements}

We wish to tank Drs. M. Aguilar, H. Bhaskara and M. Rosenbaum for discussions
and for their teachings on several subjects related to the present work. One
of us (M.S.) thanks the hospitality of the Depto. de F\' \i sica de la
Facultad de Ciencias Exactas y Naturales de la Universidad de Buenos Aires
where part of this work was performed, and Dr. D. Vergara for his comments on
BRST symmetry. We also thank the referee for pointing out several errors and
for suggestions which led to an improvement of the manuscript.

\section*{Appendix A. Cohomology of Lie algebras [12]}

Let $V$ and
${\bf g}$ respectively be a vector space and a Lie algebra over the field $k$
of real or complex numbers, and
$\phi :{\bf g}\longrightarrow End_k(V)$ a representation of
${\bf g}$ on $V$ {\it i.e.} a $k$-linear map such that
$\phi ([A,B]) = \phi (A) \circ \phi (B)  - \phi (B) \circ
 \phi (A) $. Define the vector spaces
${\cal C}^0 := V$ and for
$p \in {\bf Z}^+, {\cal C}^p := \{ t: {\bf g}\times \dots \times
{\bf g} (p$ $times) \longrightarrow V,$ {\it t multilinear totally
antisymmetric (alternating)\} }. One defines the set of linear operators
$\delta
^p:{\cal C}^p \longrightarrow {\cal C}^{p+1}, \delta ^p(t)(v_1,\ldots v_{p+1})
:= \Sigma _{i=1}^{p+1} (-1)^{i+1} \phi (v_i)(t(v_1,\ldots ,\hat v_i,\ldots
,v_{p+1}))$\\ $+ \Sigma _{1\leq i<j\leq p+1}(-1)^{i+j}
t([v_i,v_j],v_1,\ldots ,\hat v_i,\ldots ,\hat v_j,\ldots ,v_{p+1}))$ which
satisfy \\ $\delta ^{p+1} \circ \delta ^p= 0$ {\it i.e.} $\{ \delta ^p \}
_{p=0}^\infty $ is a coboundary. This leads to the complex $0 \longrightarrow
{\cal C}^0 \buildrel \delta^0 \over \longrightarrow {\cal C}^1 \buildrel
\delta^1 \over \longrightarrow {\cal C}^2 \buildrel \delta^2 \over
\longrightarrow {\cal C}^3 \buildrel \delta^3 \over \longrightarrow \cdots $
with $p$-cocycles $Z^p = ker$  $\delta ^p$ and $p$-coboundaries $B^p = im$
$\delta ^{p-1}$ , and one defines the {\it Chevalley-Eilenberg cohomology} of
${\bf g}$ with respect to the representation $\phi $ of ${\bf g}$ on $V$ (or
"with   coefficients in $V$") given by the graded group (direct sum of abelian
groups)
$H_{CE}^*({\bf g}, \phi , V; k) = \oplus _{i=0}^\infty H_{CE}^i({\bf g}, \phi
, V; k)$ with $H_{CE}^i({\bf g}, \phi , V; k) = Z^i/B^i$. In particular
$H^0_{CE} = \{ t\in V \mid t\in ker(\phi(v)), \exists  v\in {\bf g}\} =
\cap_{v\in{\bf g}} ker(\phi(v))$.

\section*{Appendix B.  Double Complexes and Total Cohomology}
{\it A double (cochain) complex} $(C, \partial, d)$ is a double array {\it
i.e.} a "lattice" of abelian groups $C^{p,q}, p,q= 0,1,2,...,$ with
differentials (group homomorphisms) $d^{p,q} :C^{p,q} \to C^{p,q+1}$ and
$\partial ^{p,q} :C^{p,q} \to C^{p+1,q}$, $d^{p,q+1} \circ d^{p,q} = \partial
^{p+1,q} \circ \partial ^{p,q} = 0$, satisfying $\partial ^{p,q+1} \circ
d^{p,q} =
d^{p+1,q} \circ \partial ^{p,q}$ {\it i.e.} the commutative diagrams

\[
\diagram{
C^{p,q}&\harr{d^{p,q}}{}&C^{p,q+1}\cr
\varr{\partial ^{p,q}}{}&&\varr{}{\partial ^{p,q+1}}\cr
C^{p+1,q}&\harr{}{d^{p+1,q}}&C^{p+1,q+1}\cr
}
\]
$(C,\partial,d)$ induces a {\it total (simple) cochain complex} $(K,D)$
as follows: for $n=0,1,2,...$ one defines the abelian groups $K^n = \oplus
_{p+q=n} C^{p,q}$ and the operators  $D^n = \oplus _{p+q=n} D^{p,q}$ with
$D^{p,q} := \partial ^{p,q} \oplus (-1)^p {d^{p,q}}$ ; then it is easy to
verify
that $D^n : K^n \to K^{n+1}$ are differentials {\it i.e.} group homomorphisms
satisfying $D^{n+1} \circ D^n = 0$, thus leading to the complex

$$K^0 \buildrel D^0 \over \longrightarrow K^1
     \buildrel D^1 \over \longrightarrow K^2
     \buildrel D^2 \over \longrightarrow \cdots $$

More pictorially\\

\vglue 4cm
\noindent where the $\circ$ dots denote the groups  $C^{p,q}$.

The cohomology of the simple complex $(K,D)$, $H^*(K,D) = \oplus_{n=0}^\infty
H^n(K,$ \\ $D)$, $H^n(K,D) = ker D^n /im D^{n-1}$ $(D^{-1} = 0)$ is called the
{\it total cohomology} of the original double complex $(C,\partial ,d)$.

A technique to compute  $H^*(K,D)$ is that of {\it spectral sequences} (SS)
[17]. A double complex has two filtrations, each having an associated SS, and
both SS's converge to the total cohomology.

\section*{Appendix C. Group Actions and Double Complexes [18]}

{\it Proposition}: Let $G$ be a Lie group, $M$ a diffentiable manifold and
$M\times G \buildrel \psi \over \longrightarrow M$ a right {\it action} of $G$
on
$M$.
Associated with this action there exists a {\it double cochain complex}
involving the Lie algebra of $G$ and the differential forms on $M$.

Proof:
{\it i}) Let $A \in {\bf g} = Lie(G)$; its fundamental vector field $A^*$ is
the vector
field on $M$ given by  $A_x^*(f) = {d\over dt}(f(xexptA))|_{t=0} = {d\over
dt}(f\circ \psi (x,exptA))|_{t=0}$ for any $x\in M$ and $f\in C^\infty
(M,{\bf R})$.

{\it ii}) If $\Phi(A^*) = \{ \phi_t \} _{t\in (-\epsilon, \epsilon)}$ is the
flow of $A^*$ then the Lie derivative of a tensor ${\tau}$ on $M$  with
respect to $A^*$ is the tensor of the same type ${\cal L}_{A^*} (\tau) =
{d\over dt} \phi _t^*(\tau) |_{t=0}$ where $\phi_t^*(\tau)$ is the pull-back
of ${\tau}$ by $\phi _t$; in particular this holds for $p$-forms on $M$ and
then for each $p=0,1,2,...,n = dim$ $M$ one has the operator ${\cal L}_{A^*}:
\Omega ^p(M) \to \Omega ^p(M)$.

{\it iii)} For each fixed $p\in \{ 0,\ldots ,n\} $ one defines the infinite
set of vector spaces:
${\cal C}_p^\nu = \{ {\bf g} \times \ldots \times {\bf g}$ $(\nu $ {\it
times)} $\buildrel \alpha \over \longrightarrow \Omega ^p(M) , \alpha $ {\it
alternating}$\} $ for $\nu \in {\bf Z}^+$ and ${\cal C}_p^0 = \Omega ^p(M) $
for $\nu = 0$.

{\it iv)}  The (canonical) representation $\phi_p : {\bf g} \longrightarrow
End( \Omega ^p(M) )$, $\phi_p (A):= {\cal L}_{A^*}$ of ${\bf g}$ on $\Omega
^p(M)$ ($\phi_p $ is a Lie algebra homomorphism since   ${\cal L}_{ [A,B] ^*}
= {\cal L}_{[A^*,B^*]} = [{\cal L}_{A^*}, {\cal L}_{B^*}]$) induces the
infinite (cochain) complex

$${\cal C}_p^0 \buildrel \delta _p^0 \over \longrightarrow {\cal C}_p^1
              \buildrel \delta _p^1 \over \longrightarrow {\cal C}_p^2
              \buildrel \delta _p^2 \over \longrightarrow \cdots $$
where $\{  \delta _p^\nu \} _{\nu =0}^\infty $ is the coboundary   $\delta
_p^\nu : {\cal C}_p^\nu \to {\cal C}_p^{\nu + 1}$, $\alpha \mapsto \delta
_p^\nu (\alpha )$: ${\bf g} \times \ldots \times {\bf g}$ $(\nu + 1$ ${\it
times)}$ $\to \Omega ^p(M)$, $\delta _p^\nu (\alpha )(A_0, \ldots ,A_\nu ) :=
\Sigma _{i=0}^\nu (-1)^i {\cal L}_{A^*_i} (\alpha (A_0,\ldots ,\hat A_i,\ldots
,$\\  $A_\nu ))$ $+ \Sigma _{1\leq i<j\leq \nu }(-1)^{i+j} \alpha
([A_i,A_j],A_0,\ldots ,\hat A_i,\ldots ,\hat A_j,\ldots ,A_\nu )$ (it holds
$\delta _p^{\nu + 1}\circ \delta _p^\nu = 0)$ which in turn induces the $p$-th
C-E cohomology of {\bf g} with coefficients in
 $\Omega ^p(M)$,  $H_{CE}^*({\bf g}, \Omega ^p(M) ) = \oplus
_{\nu=0}^\infty H_{CE}^\nu ({\bf g},$  $\Omega ^p(M) )$ with $H_{CE}^\nu ({\bf
g},
\Omega ^p(M)) = {ker \delta _p^\nu \over im \delta _p^{\nu - 1}}$.

{\it v)} Regarding the set of the previously defined C-E complexes as $n+1$
infinite (vertical) columns one defines horizontal operators $d_p^\nu : {\cal
C}_p^\nu \to {\cal C}_{p+1}^\nu $, $\alpha \mapsto d_p^\nu (\alpha ):  {\bf
g} \times \ldots \times {\bf g}$ $(\nu $ ${\it times)} \to \Omega ^{p+1}(M)$,
$d_p^\nu (\alpha )(A_1, \ldots , A_\nu )$\\ $:= d(\alpha (A_1, \ldots , A_\nu
))$   where $d$ is the De Rham operator of the manifold $M$; the commutativity
of the Lie derivative with the De Rham operator {\it i.e.}   ${\cal L}_X\circ
d = d \circ {\cal L}_X $ for all $X \in \Gamma (TM)$  implies that each
vertical ladder  $\{  d_p^\nu \} _{\nu =0}^\infty $  is a {\it cochain complex
homomorphism} and one has the following "lattice" of commuting diagrams:

\[
\diagram{
{\cal C}^0_0&\harr{d^0_0}{}&{\cal C}^0_1&\harr{d^0_1}{}&{\cal
C}^0_2&\harr{d^0_2}{}&\ldots &\harr{d^0_{n-2}}{}&{\cal C}^0_{n-
1}&\harr{d^0_{n-1}}{}&{\cal C}^0_n\cr
\varr{\delta ^0_0}{}&&\varr{\delta ^0_1}{}&&\varr{\delta
^0_2}{}&&&&\varr{\delta ^0_{n-1}}{}&&\varr{\delta ^0_n}{}\cr
{\cal C}^1_0&\harr{d^1_0}{}&{\cal C}^1_1&\harr{d^1_1}{}&{\cal
C}^1_2&\harr{d^1_2}{}&\ldots &\harr{d^1_{n-2}}{}&{\cal C}^1_{n-
1}&\harr{d^1_{n-1}}{}&{\cal C}^1_n\cr
\varr{\delta ^1_0}{}&&\varr{\delta ^1_1}{}&&\varr{\delta
^1_2}{}&&&&\varr{\delta ^1_{n-1}}{}&&\varr{\delta ^1_n}{}\cr
{\cal C}^2_0&\harr{d^2_0}{}&{\cal C}^2_1&\harr{d^2_1}{}&{\cal
C}^2_2&\harr{d^2_2}{}&\ldots &\harr{d^2_{n-2}}{}&{\cal C}^2_{n-
1}&\harr{d^2_{n-1}}{}&{\cal C}^2_n\cr
\varr{\delta ^2_0}{}&&\varr{\delta ^2_1}{}&&\varr{\delta
^2_2}{}&&&&\varr{\delta ^2_{n-1}}{}&&\varr{\delta ^2_n}{}\cr
\vdots &&\vdots &&\vdots &&&&\vdots &&\vdots \cr
}
\]

Each "square" is of the form

\[
\diagram{
{\cal C}^\nu _p&\harr{d^\nu _p}{}&{\cal C}^\nu _{p+1}\cr
\varr{\delta ^\nu _p}{}&&\varr{}{\delta ^\nu _{p+1}}\cr
{\cal C}^{\nu +1}_p&\harr{}{d^{\nu +1}_p}&{\cal C}^{\nu +1}_{p+1}\cr
}
\]

\noindent
and it holds $\delta _{p+1}^\nu \circ d_p^\nu = d_p^{\nu +1} \circ \delta
_p^\nu $.

{\it vi)} $d^2 = 0$ implies $d_{p+1}^\nu \circ d_p^\nu = 0 $ and
therefore the above lattice of abelian groups and differentials is a {\it
double complex} $({\cal C},\delta ,d)$. $\Box $

 From appendix B one has the total complex $({\cal K},{\cal D})$, with groups
${\cal K}^m = \oplus_{\nu + p = m}{\cal C}_p^\nu $ and coboundaries  ${\cal
D}^m = \oplus _{\nu + p = m} {\cal D}_p^\nu :  {\cal K}^m \to  {\cal K}^{m+1}$
for $m=0,1,2,\dots$ with  ${\cal D}_p^\nu =\delta_p^\nu \oplus (-1)^\nu
d_p^\nu$, which in turn induces the total cohomology $H^*({\cal K},{\cal D})$
of the original double complex  $({\cal C},\delta ,d)$.

\vskip1cm
\noindent
{\sc H. Garc\' \i a-Compe\' an$^1$, J.M. L\' opez-Romero$^1$, M.A. Rodr\'
\i guez-Segura}\footnote{Supported by a CONACyT Graduate Fellowship. e-mail
compean@fis.cinvestav.mx, mars@fis.cinvestav.mx}
{\it Departmento de F\' \i sica, Centro de
Investigaci\' on y Estudios Avanzados-IPN}, A.P. 14-740, 07000, M\' exico
D.F.\\[4mm]\\
{\sc  M. Socolovsky}\footnote{ e-mail
socolovs@roxanne.nuclecu.unam.mx}
\\ {\it Instituto de
Ciencias Nucleares, Universidad Nacional Aut\'onoma de M\'exico}
Circuito Exterior C.U. A.P. 70-543, 04510,M\' exico D.F.
\\[15mm]

\end{document}